# *Design flow for readout ASICs in High-energy Physics experiments*

## A. Voronin,[a] E. Malankin [b]

[a] *Skobeltsyn Institute of Nuclear Physics, Moscow State University, Moscow, 119991, Russia*

[b] *National Research Nuclear University MEPhI (Moscow Engineering Physics Institute) 115409, Kashirskoe shosse 31, Moscow, Russia*

*E-mail*: ezmalankin@mephi.ru

ABSTRACT: In the large-scale high energy physics experiments multi-channel readout application specific integrated circuits (ASICs) are widely used. The ASICs for such experiments are complicated systems, which usually include both analog and digital building blocks. The complexity and large number of channels in such ASICs require the proper methodological approach to their design. The paper represents the mixed-signal design flow of the ASICs for high energy physics. This flow was successfully implemented in the development of the readout ASIC prototypes for the muon chambers of the CBM experiment. The approach was approved in UMC CMOS MMRF 180 nm process. The design flow enables to analyze the mixed-signal system operation on the different levels: functional, behavioral, schematic and post-layout including parasitic elements. The proposed design flow allows reducing the simulation period and eliminating the functionality mismatches on the very early stage of the design.

KEYWORDS: readout ASICs; Front-end electronics for detector readout, high-energy physics.

## Contents

### 1. ASIC design approaches for HEP experiments

Mixed-signal ASIC design flow is generally described in the manuscripts [1, 2]. The design of the ASICs for high-energy physics has a lot of features. This requires a specific approach to the development process. As a peculiarity of the HEP ASIC design it could be concerned the small-batch production and low cost of the design and production. These can't be combined with the industrial scale microelectronics. Moreover, each experiment uses mostly its individual ASIC and sets the specifications to the ASIC parameters (i.e. low noise, dynamic range, high radiation tolerance, occupancy, power consumption, reliability, etc.). Currently, the ASICs for HEP are multi-channel systems, which include up to 1024 channels. This ASIC architecture limits the channel height (25 – 100 μm) and makes preferable the simple schematic solutions with the precisely adjusted parameters. Readout system includes external control of the ASIC parameters (e.g. operation modes, biasing, thresholds, channel transfer functions, etc.). The internal IC control (such as electrical calibration, monitoring of the condition, displaying the information on failures, data transfer errors, etc.) can be embedded in the chip as well. The ASICs for the space experiments have requirements which are equal to those in the space equipment. The last trend in the ASIC for HEP is the implementation of digital signal processing blocks turning the on-chip readout into the SoC. At the same time, analog processing mainly defines the quality of the ASIC full signal processing. The sophisticated architectures of this kind of ASICs require the specific complex approach to their design. The method should be built on the basis of the relevant CAD and EDA tools, by which it is possible to obtain the precise simulation results.

### 2. Design flow features

Technical specifications arise from the physics simulations (for instance by GEANT) of the expected reactions of particles (the design flow block diagram is shown in figure 1). These set up the requirements to the detector geometry, segmentation, technical realization, power, occupancy, cost and others. Then, using the detector model, it is determined the specifications of electronics (number of channels, power consumption, rate, dynamic range, etc.).

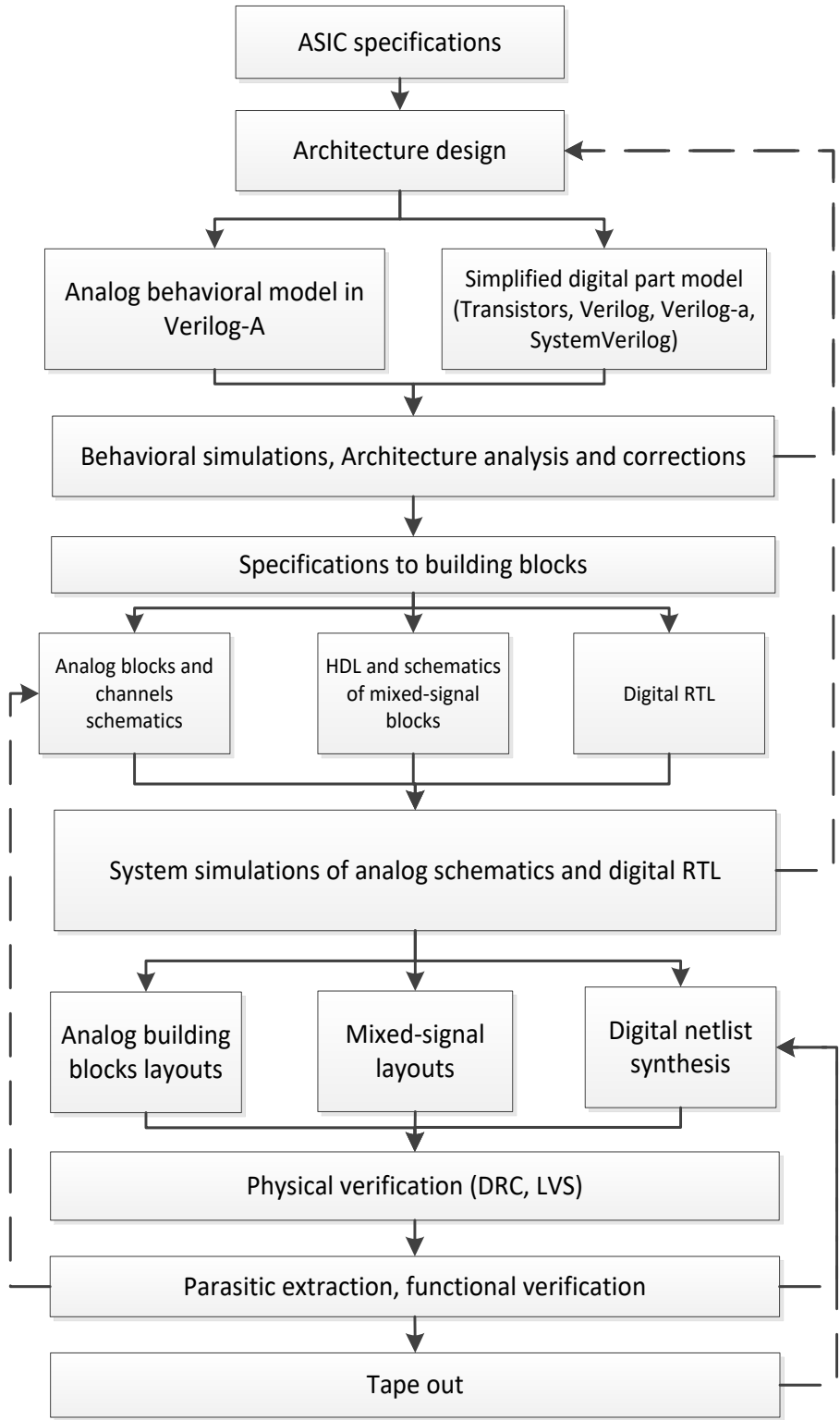

**Figure 1.** Design flow of mixed-signal ASIC for HEP and cosmic rays experiments.

On the basis of those input data it is developed one or several architecture options. All architecture ideas should be simulated on the virtual elements (for example, the behavioral HDL models) using Cadence tool set. Connection of the detector model and physical simulation of the electronics occupancy gives the correction of the requirements either the compromises to the accuracy, expenses and timelines. The first HDL models simulations set up the specifications of the building blocks. The provided studies show that the most adequate approach to the real device is given by the simulations in SPICE environment with Spectre simulator. That framework allows carrying out the mixed-signal simulation in the single environment, at the same time giving the number of parameters, taking into account the schematic non-idealities and determining the critical parameters of the schematics. Spectre permits to connect the impact generators such as noise, crosstalk at the different points, etc. Monte-Carlo SPICE simulation of the basic elements distribution determines the electronics quality both analog and digital. The ASIC building blocks designed on the transistor level or as RTL replace the correspondent behavioral part in the system model. When the behavioral block model is changed to the real one, the comparison of the parameters is carried out, as well as the simulation of mismatches in load capacity, signal polarity, time diagrams and etc.

Finally, the schematic on the transistor level (or RTL) of the whole chip is assembled. The operation of the overall system could be checked. At this stage it is possible to fix the system functionality errors and make the required corrections of the design. During the layout design besides the building blocks itself it is necessary to take into account the critical points of the schematic. Such points should have the special pads for experimental laboratory study of the prototype (e.g. with the probe station and picoprobes). It also should be provided the future laboratory tests of the separate blocks. The verification of the layouts is necessary to fix the design rules errors and the mismatches between the schematic and layout (DRC and LVS).The parasitic extraction (PEX), inclusion of the package and PCB stray elements, post layout simulations are another important step in the ASIC design. Those simulations can give the best approach of the project to the real conditions. This stage is characterized by the following types of the analysis and simulations:

- Simulations with the extracted stray elements;
- Monte-Carlo simulation of the process variations;
- Simulation of the process corners;
- Temperature variations;
- IR-drops on wires in the whole area of the die;
- Gate delays simulations;
- Integrity (signal transfer either inside the chip or on the PCB).

The flow is finalized by the extraction of the files for manufacturing. The quality of the foundry design kits and some special aspects of the simulation (analog and digital) which can give different results on the speed and accuracy should be taken into account.

## 3. Post-layout simulations of multi-channel ASICs

In the process of tuning out the behavioral model into the real (transistor or netlist) one, the simulation period can be significantly increased. In this case the optimization of simulation rate is necessary. There are several approaches which make possible to optimize the speed and accuracy of simulation:

- Utilization of the combined high-speed simulation of SPICE tools, such as APS, XPS and UltraSim. Those methods use the multi-core algorithms that give the capability to set in motion more processor cores of the calculating server. However a significant increase of digital gate in the readout ASICs a post-layout multi-core simulation doesn't give high acceleration. The operation point might be very difficult to converge. This leads to the increase of simulation time, additional iterations of simulator adjustment and incorrect results of circuit calculations.

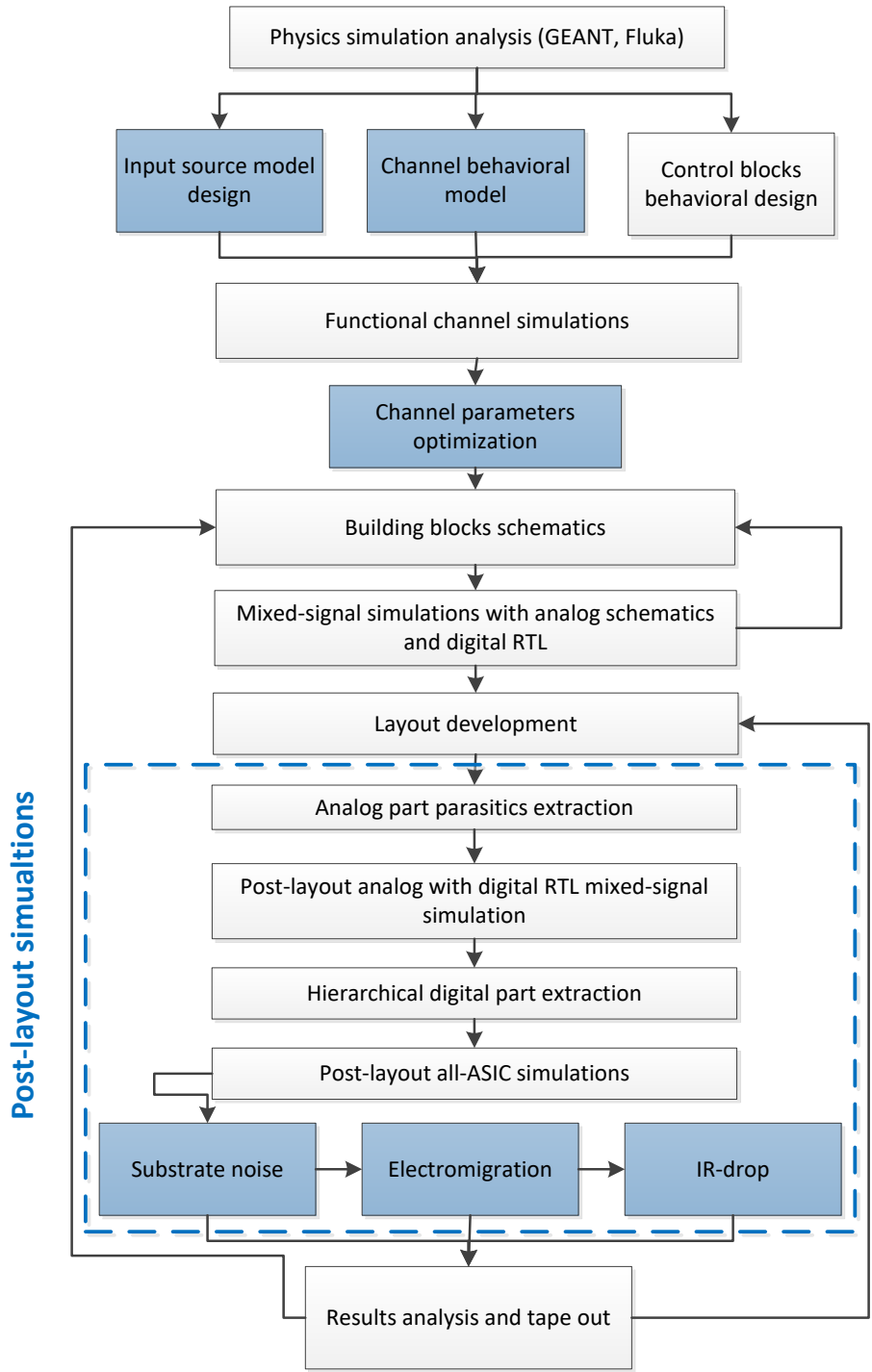

**Figure 2.** Post-layout simulation approaches in design flow for multi-channel ASICs

To provide the post-layout simulations of mixed-signal multi-channel ICs for HEP it is proposed several approaches to accelerate significantly the simulation period without losses of IC functionality evaluation accuracy:

- Hierarchical extraction. To evaluate the post-layout functionality of the multi-channel system it is not necessary to extract the stray elements inside of the digital gates. It would be enough to generate a parasitic netlist of the IC where the analog part would be represented on transistor level, but the digital part would be extracted on gate level. This shrinks the data in the netlist and significantly speeds up the simulations and operating point convergence. But it is still possible to estimate the influence of the gate switching via the signal lines and power supply buses.
- While the digital part volume increases the mixed-signal simulator (AMS) or simplified schematics solution on the high level description languages (Verilog, VHDL, Verilog-A, Verilog-AMS) are used. This method can be used to check the functionality of the system on the post-layout stage. The simulation of the analog part with stray elements take into account the influence of the crosstalk between the channel and blocks on the resulted digital processing and data transmission. This helps also to highlight the problems in analog layout, not taking into account digital stray elements.
- Substrate noise analyses, as well as Electromigration and IR-drop analyses are the new features which are offered by the up-to-date EADs. This methods utilization gives an advantage in the layout analysis of the precision, low-noise and low-power designs of multi-channel ASICs.

## 1. Methods of ASICs layout analysis

In the multi-channel ASICs for HEP the digital gates switching can negatively influence on the analog blocks and system performance. If the magnitude of the digital crosstalk is of the analog channel noise level the dynamic range, gain and bandwidth can decrease dramatically. In some cases the crosstalk can cause oscillations [4]. Use of the submicron CMOS processes leads to a decrease of power supply voltages. High functionality of ASICs requires the control of the supply currents and reliability anticipation of the IC before manufacturing.

There are several methods which can be applied to analyze the multi-channel layout:
- Parasitic netlist extraction,
- Analysis of the noise over silicon substrate,
- power supply IR-drop analysis,
- Electromigration in signal and power supply metal lines analysis.

The coupling mechanisms of the noise signal to substrate describes in details in [5 – 8]. There are plenty of methods to minimize the crosstalk in mixed-signal ICs:
- Reduction of digital gates generating the noise while switching,
- Analog design with digital noise immunity (e.g. differential and low-power circuits),
- Electromagnetic isolation of analog and digital part,
- Increase of the distance between the source of the noise and «victim»
- Shielding by different layers

Currently the following simulators created to analyze the substrate noise: SubstrateStorm, HFSS, SeismIC, MEDICI, SPACE, Sequoia [6].

To simulate the substrate noise it is necessary to have a macromodel which is generated on the bases of the layers and doping information. The flow to generate the substrate macromodel is described in [9, 10]. To analyze the substrate noise it is necessary to pass DRC and LVS checks and then extract the parasitic netlist using the substrate macromodel. This type of extraction allows AC simulation with substrate noise and displaying the results onto the layout view (see Figure 3).

After the substrate stray elements have been extracted the simulation of the substrate noise can be successfully implemented in the multi-channel readout ASIC design. The most sensitive parts of the ASIC are CSA and shaper. To analyze the influence of the substrate noise one should pass the following steps:
- Substrate extraction
- Simulation of CSA + shaper channel with substrate (see figures)
- Study of the noise amplitude at sensitive points vs the distance to noise source point.
- Study of the dependence of the noise on the digital signal magnitude
- Analyze of the distribution paths of substrate noise
- Determination of the optimum location of the input CSA transistor
- Analyzes of electromagnetic shielding (n-well, triple-wells, guard rings etc.)

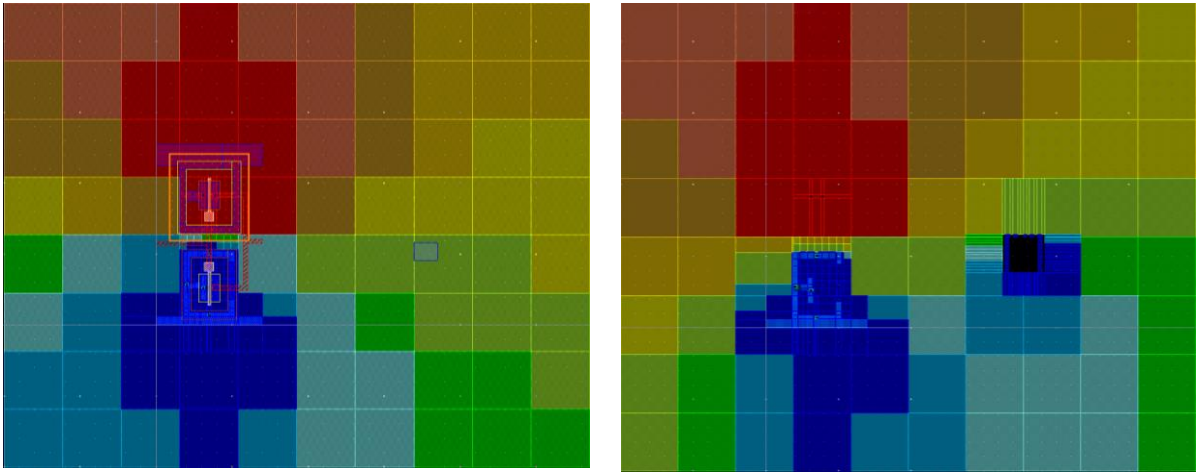

**Figure 3** Substrate noise simulation

The substrate noise study shows for 180 nm process [7] that application of n-well simultaneously with p+ - guard ring decreases the S21 parameter of substrate noise distribution on 40 dB in 1 GHz bandwidth. The n-well with p+ - guard ring around the analog part was implemented in the design of the readout multichannel ASIC for CBM experiment [3] (see figure 4).

IR-drops can be the crucial point in the multi-channel readout ASIC design for HEP. The IR-drops causes by non-zero resistance of the IC metallization and proportional to the current. The effect can be observed both in dynamic and in static. While the power supply voltage in the up-to-date ASIC decreased the effect can be more substantial. This effect can be simulated with the standard extraction netlist using the tools like Cadence Assura QRC and PVS, Mentor Graphics PEX.

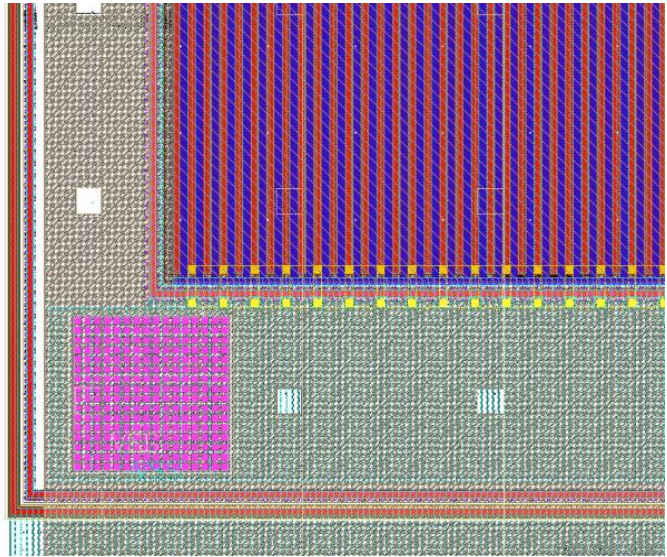

**Figure 4** Fragment of the CSA layout with n-well and p+ - guard ring

But this approach doesn't make sense because the multi-channel ASIC area is usually about several mm$^2$ and layout contains a lot of transistor and digital gates. This dramatically increases the volume of data to simulate and slowing down the simulation process and failing the simulator.

The solution is to use the approach based on the DSPF-netlist extraction for specified nets and subsequent simulations with Voltus-Fi tool [11] by Cadence (see figure 5).

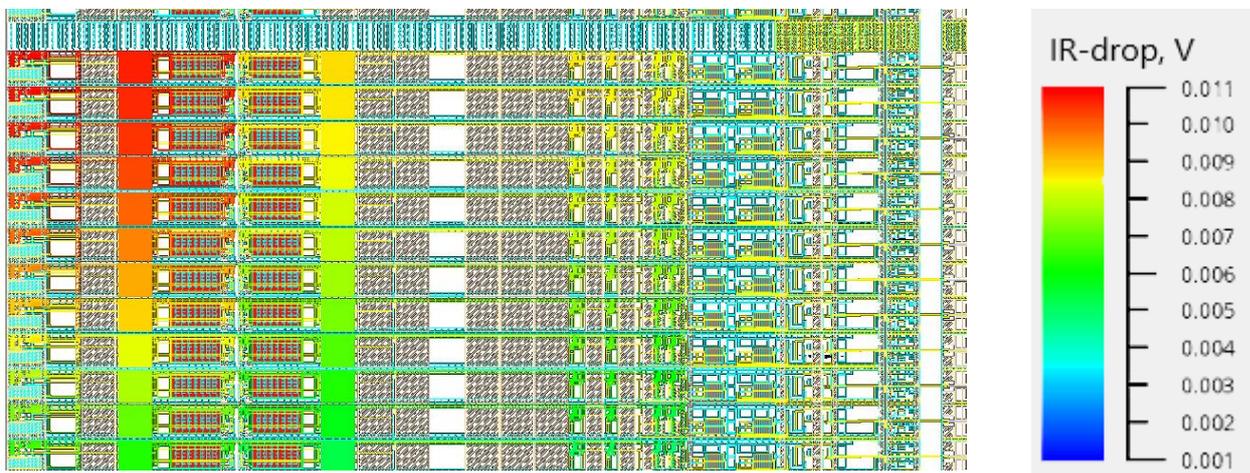

**Figure 5** IR-drop simulation of the ASIC for CBM MUCH

The reliability of the readout ASICs is a key point in the design. Improper sized nets lead to a breakdown of the line due to the electromigration[13]. To control current densities and adjustment of the geometry can be realized using the dspf-netlist as well and electromigration analysis. (see. fig. 6).

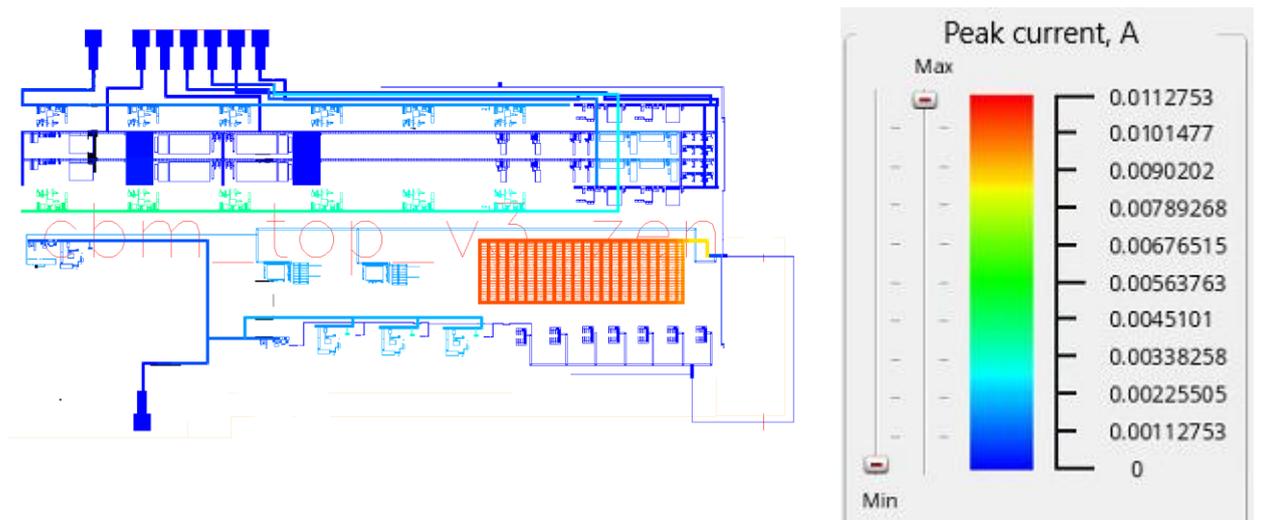

**Figure 6** EM simulation of the ASIC for MUCH

## 4. Conclusions

The readout ASIC design flow for high energy physics applications was described. The design flow makes possible to analyze the mixed-signal system operation on the different levels: functional, behavioral, schematic and post layout including parasitic elements. The proposed design flow allows reducing the simulation period and eliminating the functionality mismatches on the very early stage of the design. It was successfully embedded to the development of the read-out ASIC prototype for the muon chambers of the CBM experiment. The approach was approved in UMC CMOS MMRF 180 nm process.